\documentclass[prc,aps,twocolumn,floats,floatfix,showpacs,superscriptaddress,nofootinbib]{revtex4}
\usepackage{graphicx}

\newcommand{\nuc}[2]{$^{#1}${#2}}
\newcommand{\etal}{\emph{et al.}}

\begin{document}
\title{Beyond mean-field study of excited states:
       Analysis within the Lipkin model}

\author{A. P. Severyukhin}
\affiliation{Bogoliubov Laboratory of Theoretical Physics,
             Joint Institute for Nuclear Research,
             141980 Dubna, Moscow region, Russia}
\affiliation{PNTPM, CP229,
             Universit{\'e} Libre de Bruxelles,
             B-1050 Brussels, Belgium}

\author{M. Bender}
\affiliation{NSCL, Michigan State University, 
             East Lansing, MI 48824, USA}
\affiliation{L'Espace de Structure Nucl{\'e}aire Th{\'e}orique,
             Dapnia/SPhN, CEA Saclay, 91191 Gif sur Yvette Cedex, 
             France}

\author{P.-H. Heenen}
\affiliation{PNTPM, CP229,
             Universit{\'e} Libre de Bruxelles,
             B-1050 Brussels, Belgium}

\begin{abstract}
Beyond mean-field methods based on restoration of symmetries and
configuration mixing by the generator coordinate method (GCM)
enable to calculate on the same footing correlations in the ground
state and the properties of excited states. Excitation energies
are often largely overestimated, especially in nuclei close to
magicity, even when transition probabilities are well-described.
We analyse here the origin of this failure. The first part
of the paper compares realistic projected GCM and QRPA
calculations for selected Sn isotopes performed with the same
effective Skyrme interaction. Although it is difficult to perform
RPA and GCM calculations under exactly the same conditions, this
comparison shows that the projected GCM overestimates the RPA
results. In the second part of this paper, we compare GCM and RPA
in the framework of the exactly solvable Lipkin-Meshkov-Glick
model. We show that the discretized GCM works quite well and
permits to obtain nearly exact results with a small number of
discretization points. This analysis indicates also that to break
more symmetries of the nuclear Hamiltonian in the construction of
the GCM basis is probably the best way to improve the description
of excited states by the GCM.
\end{abstract}


\date{March 28 2006}

\maketitle

\section{Introduction}
Self-consistent mean-field methods are one of the standard
microscopic approaches in nuclear structure theory \cite{RMP}.
At present, they are the only available microscopic method
that can be systematically applied on a large scale for medium and heavy
nuclei. The full model space of occupied states can be used, which
removes any distinction between core and valence particles and the
need for effective charges. This allows the use of a universal effective
interaction, universal in the sense that it can be applied for all
nuclei throughout the periodic chart.

Despite its successes, the self-consistent mean-field method has a
number of well-known limitations. From a conceptual point of view,
the mean-field approach is designed to describe mainly
ground-state properties, and gives very limited access to excited
states. This is in contrast to the microscopic methods that are
available for light nuclei, like the no-core shell model and the
shell-model Monte-Carlo, which also describe excitation spectra. A
systematic way to resolve these problems is offered by symmetry
restoration and configuration mixing. Several groups now develop
methods going beyond a mean-field approach based on the generator
coordinate method (GCM), either with non-relativistic Skyrme
\cite{Val00a,BH03} or Gogny \cite{Rod00a} interactions, or with
relativistic Lagrangians \cite{Nik06a}, see also
\cite{Egi04a,enam} and references given therein. The aim is to
obtain a global description of ground states, including collective
correlations which cannot be included in a mean-field approach,
even at the effective interaction level, and of excited states of
all nuclei in a single, unified method.

An introduction to the method that we develop along these lines
can be found in Ref.~\cite{enam}. First applications have
demonstrated that such a method permits to describe the energies
of low-lying collective excitations and electric transition
probabilities, in-band and out-of-bands. However, in many cases it
has also been found that whenever states can be grouped into
rotational bands, the spectra obtained with the GCM are too spread
and that the excitation energies are too high. For spherical
nuclei, in particular those close to doubly-magic ones, the
low-energy collective spectra are only qualitatively in agreement
with the data. This feature is well illustrated by a detailed
study of the quadrupole and octupole modes of nuclei around
\nuc{208}{Pb} in Ref.\ \cite{Hee01}: the excitation energies of
the first $3^-$ and $2^+$ states are overestimated by more than
1~MeV. On the other hand, the energies of giant resonances were
found to be more realistic, as was also shown in Ref.\
\cite{SH96}.

The origin of this problem is not obvious and it cannot be
expected to be unique. A possible source of error could be the
inadequacies of effective interactions to describe spectra. It is
certainly an appealing feature of the projected GCM that the same
effective interaction can be used to generate the mean-field
states and to perform their mixing. However, existing interactions
are still far from describing all nuclei with a similar high
quality~\cite{Gor02a,Ber05a,Ben06a}. Furthermore, they have been
adjusted exclusively at the level of the mean-field approximation
and on ground state properties, with at most constraints on the
values of global parameters, like the effective masses or the
compressibility, deduced from the systematics of excited states.
There is no guarantee that such effective interactions will
correctly predict spectra, although, up to now, GCM results are
encouraging, and, in most cases, in good qualitative agreement
with available data. Another source of uncertainty of the GCM
comes from the choice of the variational space in which the
configuration mixing is performed. It is usually constructed by
introducing constraints on one or a few collective variables related
to the shape of the nuclear density in the mean-field equations.
Such a choice may be more appropriate to describe the
properties of ground states than those of excited states, for
which additional degrees of freedom might have to be included.

Up to now, these problems have not been addressed in a systematic
way, with the exception of the detailed study of \nuc{208}{Pb} and
its isotopes mentioned above but which did not include the
restoration of rotational symmetry. Rather than to repeat the same
kind of analysis as in Ref.~\cite{Hee01}, we will adopt here
another strategy. We will first compare results obtained with the
same effective interaction using either the GCM or the random
phase approximation (RPA). The RPA is an alternative microscopic
method to calculate collective excitations, relying also on the
existence of a nuclear mean field. The RPA is very well adapted to
the description of collective excitations in spherical nuclei.
This will enable us to separate the problems related to the
effective interactions from those related to the method. We will
make this comparison for two Sn isotopes, the doubly magic
\nuc{132}{Sn} and \nuc{120}{Sn}, performing the GCM and RPA
calculations under conditions as close as possible. We will see
that the results differ in a manner that raises questions about
the degrees of freedom that are explicitly included in the GCM.
However, significant differences between both methods cannot be
easily eliminated and the comparison between both models cannot be
fully conclusive.

One therefore needs a view on the problem from another
perspective. Exactly solvable models constitute a very fruitful
ground for the test of and comparison between many-body methods. They
also permit to explore new developments at a very limited cost.
For this purpose, we need a model where collective variables
similar to deformations can be introduced and where a discretized
version of the GCM can be defined in a way similar to that of the
realistic applications. The Lipkin-Meshkov-Glick (LMG) model,
introduced in Ref.~\cite{LMG65}, has the required properties.
Depending upon the strength of the interaction, two different
kinds of solutions are obtained at the Hartree-Fock (HF)
level~\cite{RingSchuck}: "spherical" ones at low values of the
strength, and  "deformed" ones beyond a critical strength. The
second part of the paper is devoted to a detailed discussion of
the LMG model.

%
%
\section{Comparison between the QRPA and projected GCM on Sn isotopes}

\subsection{Technical aspects}

Our beyond mean-field method has already been presented in details
and applied to a large number of
nuclei~\cite{Val00a,BH03,BFH03,DBBH03,BHB04,BBH05,BBH06}. Let us here
summarize those of its features which are essential for a critical
comparison with the RPA.

The starting point of the method is a set of constrained
mean-field calculations, with the axial quadrupole moment as a
constraining operator. The wave functions obtained for a discrete
set of quadrupole moments are projected on good particle numbers
and on angular momentum. For each value of the angular momentum,
the projected wave functions are mixed with respect to the
quadrupole moment by the generator coordinate method (GCM),
leading for each angular momentum to a collective wave function
spread over a range of deformations.

The same effective interactions, the Skyrme SLy4 \cite{sly4} in
the mean-field (particle-hole) channel and a density-dependent
zero-range force \cite{pairing} in the pairing (particle-particle) channel
with a strength of $-1250$~MeV~fm$^{3}$ are used for the construction
of the mean-field wave functions and for the calculation of the
GCM matrix elements. The BCS subspace is limited to an energy
range of 5~MeV above and below the Fermi level.

Pairing correlations are a necessary ingredient of a GCM
calculation: the Hartree-Fock Slater determinants corresponding to
two deformations, for which the number of occupied single-particle
states of a given symmetry is different, are orthogonal. This
feature makes a GCM calculation numerically unstable. The problem
is cured by the partial occupation of single-particle levels due
to pairing. As BCS pairing correlations collapse whenever the
density of single-particle levels around the Fermi surface is low,
we use the Lipkin-Nogami (LN) method to ensure that pairing
correlations are present for all values of the quadrupole moment.

\begin{figure}[t!]
\includegraphics[width=7cm]{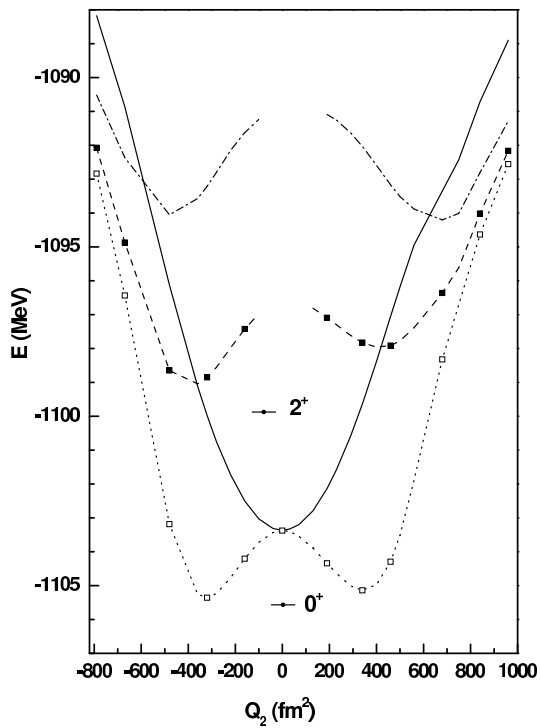}
\caption{\label{E_Sn132}
Projected energy curves as a function of the quadrupole moment for
\nuc{132}{Sn}. The particle-number projected mean-field
deformation energy is represented by a solid line, while dotted,
dashed and dash-dotted lines correspond to projected mean-field
energies for \mbox{$J=0$}, 2, and 4, respectively.}
\end{figure}

%
%
\subsection{Results for Sn isotopes}

Let us first compare GCM and RPA predictions for the first excited
$2^+$ state in a doubly magic nucleus, \nuc{132}{Sn}, which has
already been extensively studied by the
RPA~\cite{TENS02,colo03,giam03}.

\begin{figure}[b!]
\includegraphics[width=8cm]{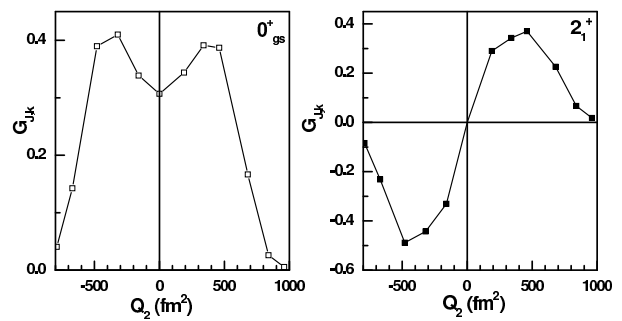}
\caption{\label{Wf_sn132}
Collective GCM wave functions for the ground state and the first
$2^+$ state of \nuc{132}{Sn}.}
\end{figure}

In Fig.\ \ref{E_Sn132}, the mean-field energy curve is shown
together with the \mbox{$J=0$} to 4 projected energy curves as a
function of the quadrupole moment. As usual, the projected states
are labelled by the deformation of the mean-field state which is
projected. The topography of the projected curves is typical for
nuclei which have a well defined spherical mean-field ground
state~\cite{BBH06}. After projection, there are two nearly
degenerate \mbox{$J=0$} minima corresponding to two slightly
deformed mean-field configurations, oblate and prolate. The energy
gain is moderate, of the order of 2~MeV, with a marginal extra
gain of 210 keV due to configuration mixing. The collective GCM
wave functions are plotted in Fig.~\ref{Wf_sn132} for \mbox{$J=0$}
and 2.

In Table~\ref{be2_132}, we compare the energies and $B(E2)$ values
obtained in projected GCM and RPA calculations based on the same
effective interaction~\cite{colo03,colo04} with the experimental
data~\cite{ram01,rad05}. The excitation energy of the $2^+$ state
is overestimated by both methods. Within the RPA framework, one can
show that, for Sn isotopes, the inclusion of phonon couplings
decreases the excitation energy of the $2^+$ state and reduces the
$B(E2)$ transition probability, although this decrease is less
pronounced in \nuc{132}{Sn} than in lighter isotopes \cite{SVG04}.
Note also that the energy centroid of the isoscalar giant
quadrupole resonance is predicted correctly by the GCM at
12.45~MeV, close to the value given by the empirical systematics
$63$ $A^{-1/3}$ MeV.

We have performed a similar calculation for \nuc{120}{Sn} for
which neutron pairing correlations are present in the ground state
and for all deformations. Results are compared in
Table~\ref{be2_120} with a RPA calculation~\cite{colo04} and with
experiment~\cite{ram01}. Again, we find a discrepancy between the
RPA and GCM predictions for the excitation energy of the low-lying
$2^+$ state, whose energy is significantly larger in the GCM
calculation than in RPA. The $B(E2)$ value obtained within GCM is
closer to the experimental value than the RPA value. This result
is consistent with our findings for other systems that transition
moments are in most cases much better described than excitation energies,
and suggests that also for nuclei that are spherical at the mean-field
level the geometrical properties of excited states are better described
by the GCM than their excitation energies. This may be related to
the fact that the components of the $2^+$ states corresponding
to two-quasi-particle excitations breaking time reversal invariance
are completely missing from the GCM model space and that these
components have a larger contribution to energies than to
transition probabilities.

\begin{table}[t!]
\caption{\label{be2_132}
Excitaion energy (in MeV) and $B(E2)$ value (in e$^2$ fm$^4$) for
up transition to the first $2^{+}$ state in \nuc{132}{Sn}.
}
\begin{tabular}{lcc}
\hline \noalign{\smallskip}
   Method                   & Energy (MeV)& B(E2$\uparrow$) \\
\noalign{\smallskip} \hline \noalign{\smallskip}
GCM                         &5.69           &630                             \\
Particle-hole RPA           &5.13           &1370                            \\
Experiment                  &4.04           &1100$\pm$300                    \\
\noalign{\smallskip} \hline
\end{tabular}
\end{table}

\begin{table}[t!]
\caption{\label{be2_120}
Excitation energy (in MeV) and $B(E2)$ value (in e$^2$ fm$^4$) for
up transition to the first $2^+$ state in \nuc{120}{Sn}
}
\begin{tabular}{lcc}
\hline \noalign{\smallskip}
   Method                   & {Energy (MeV)}& {B(E2$\uparrow$) }\\
\noalign{\smallskip} \hline \noalign{\smallskip}
GCM                         & 2.40           & 1350            \\
Quasiparticle RPA           & 1.44           & 440            \\
Experiment                  & 1.17           & 2020$\pm$40    \\
\noalign{\smallskip} \hline
\end{tabular}
\end{table}

Since neither RPA nor GCM do satisfactorily reproduce the
experimental energies and transition probabilities for both Sn
isotopes, it is tempting to conclude that the Skyrme effective
interaction is not fully adequate to describe low energy
excitations in this mass region.

Besides this problem of interaction, a direct comparison between
RPA and GCM results shows a discrepancy and a suspicion that the
GCM variational subspace as defined in actual applications is not
as rich as the RPA one, in particular, at small deformations, where
two-quasi-particle excitations breaking time-reversal invariance are
not included in our GCM. Unfortunately, there are several differences
between both calculations, which might affect their comparison,
some of them being hard to eliminate.

In both cases, the same effective interaction is used in the
mean-field channel. However, since \nuc{132}{Sn} is a doubly-magic
nucleus, there are no pairing correlations in the spherical ground
state with either a BCS or a Bogoliubov treatment of pairing.
Therefore, there is no pairing at all in an RPA approach, in
contrast to the GCM. As soon as the quadrupole deformation of
the constrained mean-field state is sufficiently large (in this case,
500 fm$^2$), pairing correlations are present also in the BCS method.
As already mentioned above, the GCM requires generating wave functions
which vary smoothly along the collective path, which is enforced
using of the LN prescription; hence, in our GCM pairing correlations
are present in all states, even in the spherical mean-field
configuration of \nuc{132}{Sn}.

Another difference between the RPA calculation of \cite{colo04} and
our GCM calculation is that the Coulomb and the spin-orbit residual
interactions are not included
in the RPA, but are always present in the GCM. The effect of these
terms has been very recently studied by Terasaki \etal\
\cite{ter05} and P{\'e}ru \etal\ \cite{Per05a}. 
The inclusion of the Coulomb residual interaction
raises the $2^+$ excitation energy by 200 to 300~keV.

Some recent studies have also pointed out that the RPA has shortcomings
which are usually forgotten. As shown by Johnson and Stetcu, the QRPA
does not restore symmetries exactly \cite{Ste03a}. Moreover, in
some cases, the RPA predicts poorly the ground state correlation
energies \cite{Ste02a}, the excitation energies \cite{Joh02a} and
$B(E2)$ values \cite{Ste03a}.

Still, it seems clear that the GCM overestimates excitation
energies in spherical nuclei close to magicity.  On the other
hand, since it looks difficult to remove all the differences
between GCM and RPA calculations, the use of a model that can be
exactly solved seems to be the most appropriate way to deepen the
present analysis.
%
%
\section{Excited states in the Lipkin-Meshkov-Glick model}

\subsection{The model}

Lipkin, Meshkov, and Glick introduced an exactly solvable model
\cite{LMG65}, usually called "Lipkin model" or "LMG model" in
the literature, that has been widely used to test methods of
approximation for the nuclear many-body problem.

The model consists of $N$ fermions distributed in two $N$-fold
degenerate shells separated by an energy $\varepsilon$. In their
original paper, two different Hamiltonians were proposed. The one
which is the most usually studied contains a monopole-monopole
interaction and is given by:
\begin{equation}
\label{hex}
\hat{H}
= \varepsilon J_{0}
  - \frac{1}{2} V ( \hat{J}_{+} \hat{J}_{+} + \hat{J}_{-} \hat{J}_{-} )
,
\end{equation}
where $V$ is the interaction strength and $\hat{J}_{0}$, $\hat{J}_{\pm}$
are quasi-spin operators \cite{LMG65,ho73,RingSchuck}
\begin{eqnarray}
\hat{J}_{0}
& = & \frac{1}{2}\sum_{p=1}^{N}
      (   \hat{c}_{+p}^{\dagger} \hat{c}_{+p}
        - \hat{c}_{-p}^{\dagger} \hat{c}_{-p} )
      ,
      \nonumber \\
\hat{J}_{+}
& = & \sum_{p=1}^{N} \hat{c}_{+p}^{\dagger} \hat{c}_{-p}
      ,
      \nonumber \\
\hat{J}_{-}
& = & \hat{J}_{+}^{\dagger}
      .
\end{eqnarray}
with the algebra
\begin{equation}
[\hat{J}_{+},\hat{J}_{-}]
= 2 \hat{J}_{0}
, \qquad
[\hat{J}_{0},\hat{J}_{\pm}]
= \pm \hat{J}_{\pm}
.
\end{equation}
The operators $\hat{c}_{+p}^{\dagger}$ and
$\hat{c}_{-p}^{\dagger}$ create a particle in the upper or lower
shells, respectively, where $p$ labels the $N$ degenerate levels
within the shells. The operator $\hat{J}_0$ measures half of the
difference between the number of particles in the upper and the
lower levels.

The exact wave functions are eigenstates of two operators, the
total quasispin operator $\hat{J}^2 = \frac{1}{2} (\hat{J}_{+}
\hat{J}_{-} + \hat{J}_{-} \hat{J}_{+} ) + \hat{J}_0^2$ with
eigenvalue $J (J+1)$, and a signature operator $e^{i \pi
\hat{J}_0}$, which, for an even number of particles, has two
eigenvalues equal to $\pm 1$. Therefore, as discussed in detail in Ref.\
\cite{LMG65}, the interaction does not mix states which have
different eigenvalues of $\hat{J}^2$ and $e^{i \pi \hat{J}_0}$ and
the Hamiltonian matrix splits into blocks, which are multiplets
in $J$ of order $2J+1$. The multiplets separate further into
blocks of size of $J$ and $J+1$ corresponding to the two values
for the signature.

To understand the connection between the LMG model and realistic
nuclear models, it is interesting to identify the structure of its
eigenstates~\cite{WN}. In the limit of vanishing interaction
strength \mbox{$V = 0$}, the exact ground state corresponds to an
independent particle state with all the lower single-particle
levels occupied, while the exact first excited state is given by a
1p-1h excitation on top of the ground state, the second excited
state by 2p-2h excitations, etc.

As the Hamiltonian does not mix states which different $J$ values,
the exact wave functions are linear combinations of the
\mbox{$2J+1$} eigenfunctions of the operator $\hat{J}_{0}$ within
a multiplet of given $J$ \cite{LMG65}. There is one state of each
possible $n$p-$n$h content in each multiplet. The non-interacting
ground state has \mbox{$J_0 = -N/2$}, and belongs to the multiplet
with maximum $J$, i.e.\ \mbox{$J = N/2$}. The first
non-interacting excited state is a 1p-1h state which has
\mbox{$J_0 = -N/2+1$}. Pure 1p-1h states are admixtures of states
from the \mbox{$J = N/2$} and \mbox{$J = N/2-1$} multiplets, while
pure 2p-2h states are admixtures of states within the \mbox{$J =
N/2$}, \mbox{$J = N/2-1$} and \mbox{$J = N/2-2$} multiplets, etc
for higher $n$p-$n$h excitations until \mbox{$n = N$}.

For small values of the interaction strength $V$, the mixing within
the multiplets should be small and  low-lying levels should have
a similar structure as the non-interacting ones. By contrast, for large values
of $V$, the eigenstates will exhibit a complicate mixing of many p-h
excitations. Thus, the model exhibits a transition between shell-model-like
states and collective states.

%
%
\subsection{Mean-Field Approximation}

In mean-field, or Hartree-Fock (HF), approximation, the many-body wave
function $| \alpha, \varphi \rangle$ is given by a Slater determinant
\begin{equation}
| \alpha, \varphi \rangle
= \prod_{p=1}^{N} \hat{a}_{0p}^{\dagger} | - \rangle
,
\end{equation}
characterized by two real degrees of freedom $\alpha$ and $\varphi$, that
will be specified below.
The particle- and hole-creation operators of the corresponding
HF single particle basis are given by a unitary transformation among the
operators corresponding to the non-interacting basis \cite{RingSchuck,ho73}:
\begin{equation}
\left( \begin{array}{c}
       \hat{a}_{1p}^{\dagger} \\
       \hat{a}_{0p}^{\dagger}
       \end{array}
\right)
= \left( \begin{array}{cc}
         \cos (\alpha )                 & -\sin (\alpha ) \, e^{-i\varphi} \\
         \sin (\alpha ) \, e^{i\varphi} &  \cos (\alpha )
         \end{array}
  \right)
  \left( \begin{array}{l}
         \hat{c}_{+p}^{\dagger} \\
         \hat{c}_{-p}^{\dagger}
         \end{array}
  \right)
.
\end{equation}
The subscripts 0 and 1 denote hole and particle states,
respectively. The variables $\alpha$ and $\varphi$ vary both in
the interval $[-\pi/2,\pi/2]$. They can be identified as
constraints, that, due to the low dimensionality of the LMG model,
map the entire space of mean-field states. New quasi-spin
operators corresponding to the states with finite $\alpha$ and
$\varphi$ can be easily constructed, as described, for example, in
\cite{ho73}. An alternative manner to write the constrained HF
states will be useful in the context of the GCM. Using the
Thouless theorem \cite{T60}, the normalized constrained HF states
$|\alpha ,\varphi \rangle$ can be obtained from the
non-interacting ground state, that corresponds to \mbox{$\alpha =
\varphi = 0$}, as
\begin{equation}
\label{HF:thouless}
|\alpha ,\varphi \rangle
= \cos^N ( \alpha )
  \exp \left[ \tan (\alpha)  \exp (i\varphi) \hat{J}_{+} \right]
  | \alpha = 0, \varphi = 0 \rangle
.
\end{equation}
A pointed out by Bhaumik \etal\ \cite{Bha81a}, the constrained HF
states of the LMG model can also be formulated in the language
of coherent states, which allows to make use of generating functional
techniques to calculate matrix elements \cite{Zha90a}.

The constrained HF ground-state energy is a function of the
variables $\alpha$ and $\varphi$:
\begin{equation}
E_{gs}^{\text{HF}}(\alpha ,\varphi )
= -\frac{\varepsilon N}{2}
   \left[ \cos (2\alpha)
          + \frac{1}{2} \chi \sin^{2} (2\alpha) \cos (2\varphi)
   \right]
,
\end{equation}
where
\begin{equation}
\chi
= \frac{\left( N-1\right) V}{\varepsilon}
.
\end{equation}
Note that, for a given value of $\alpha$, the lowest HF state
always corresponds to \mbox{$\varphi = 0$}. The eigenvalues
of the single-particle Hamiltonian, usually called single-particle
energies, depend on $\alpha$ only for any mean-field state
$|\alpha ,\varphi \rangle$.

One can identify the variable $\alpha$ as a deformation parameter.
There is a phase transition at \mbox{$\chi = 1$} from a spherical
(\mbox{$\alpha = \varphi = 0$}) to a "deformed" ground state. In
the latter case, the value of $\alpha$ is obtained by solving the
equation \mbox{$\chi \cos (2\alpha) = 1$}. The phase transition
and the properties of exact and approximated ground states in this
regime were first discussed by Agassi \etal\ \cite{Aga66a}.

While the HF states remain eigenstates of $\hat{J}^2$, "deformed"
HF states break the signature symmetry of the exact solutions
(which is often called "parity" in the literature) for any
non-zero value of $\alpha$. The HF states mix the $n$p-$n$h states
with even and odd $n$ within a given $J$ multiplet. As a
consequence, the constrained HF states for non-zero interaction
strength contain 0p-0h, 1p-1h, 2p-2h, 3p-3h etc states. The
$n$p-$n$h components with even and odd $n$ can be separated with
a projection operator \cite{Aga66a,Rob92a}. Due to the simple
structure of the LMG model with one relevant coordinate only,
minimization of the energy obtained by projection after variation
is equivalent to projection before variation \cite{Rob92a}.

The signature symmetry is a discrete symmetry, in contrast to the
continuous rotational symmetry broken in nuclei with a quadrupole
deformation. Li \etal\ \cite{Li70a} have introduced a
generalization of the LMG model that can been used to test
techniques for approximate angular-momentum projection, see
\cite{Hag00a} and references therein. The structure of the
original LMG model is closer to parity projection in
octupole-deformed nuclei \cite{Rob92a}.

The interpretation of the $\varphi$ degree of freedom is less intuitive.
It enters the HF states as a phase. It is explored by time-dependent HF
(TDHF) states, hence a necessary ingredient of any dynamical model
\cite{KLDG80,Kri77a}. Using the variables $\alpha$ and $\varphi$,
one can form a set of
two canonically conjugate variables with which, for example, the
time-dependent HF equations can be transformed to classical
equations of motion~\cite{KLDG80}.

%
%
\subsection{Random phase approximation}
The RPA of the LMG model was formulated for the first time in
Ref.~\cite{LMG65}. The RPA is usually constructed on top of the
"spherical" HF state $| \alpha = 0, \varphi = 0 \rangle$, which is
the ground state for \mbox{$\chi < 1$}. In this regime, the RPA
phonon creation operator, defined as a superposition of all
possible 1p-1h excitations, is given by
\begin{equation}
\hat{Q}^{\dagger}
= \frac{1}{\sqrt{N}} ( X \hat{J}_{+} - Y \hat{J}_{-} )
.
\end{equation}
One assumes that the ground state is the RPA phonon vacuum
$| 0 \rangle $, i.e.\ \mbox{$\hat{Q} | 0 \rangle = 0$}.
The first excited state is given by $\hat{Q}^{\dagger} | 0 \rangle$
with the normalization condition:
\begin{equation}
\langle 0 | [ \hat{Q} , \hat{Q}^{\dagger} ] | 0 \rangle
= X^{2} - Y^{2}
= 1
.
\end{equation}
Profiting from the simplicity of the LMG model, the authors of
\cite{LMG65} have solved \mbox{$\hat{Q} | 0 \rangle = 0$} exactly.
The usual way is to solve the RPA equations in the space of 1p-1h
excitations by linearization. Making use of the equation-of-motion
approach~\cite{Row68a,R70}:
\begin{equation}
\langle 0 | [ \delta \hat{Q}, [ \hat{H},\hat{Q}^{\dagger} ] ] | 0 \rangle
= (E-E_0) \langle 0 | [ \delta \hat{Q},\hat{Q}^{\dagger} ] | 0 \rangle
,
\end{equation}
where $E$ is the absoute energy of the RPA state, and $E_0$ the energy of
the RPA ground state, one obtains the RPA equations:
\begin{equation}
\left(
\begin{array}{cc}
 \mathcal{A} & \mathcal{B} \\
-\mathcal{B} & -\mathcal{A}
\end{array}
\right)
\left(
\begin{array}{c}
X \\
Y
\end{array}
\right)
= (E-E_0) \,
\left(
\begin{array}{c}
X  \\
Y
\end{array}
\right),
\end{equation}
where \mbox{$\mathcal{A} = \varepsilon$} and
\mbox{$\mathcal{B} = - ( N - 1 ) \, V = - \varepsilon \chi$}.

From these equations, the excitation energy of the first excited state of
the Hamiltonian~(\ref{hex}) within the RPA is found to be
\begin{equation}
E-E_0
= \varepsilon \sqrt{1 - \chi ^{2}}
.
\end{equation}
This energy is equal to zero for \mbox{$\chi=1$}, where the system
undergoes a phase transition, and becomes imaginary for even larger
values of $\chi$.

The RPA is explicitly constructed as a superposition of 1p-1h states;
hence, it automatically contains the right physics of the lowest
excited state in the limit \mbox{$\chi \to 0$} and should be accurate in
the limit of small $\chi$.

The RPA correlation energy~\cite{RingSchuck} in the ground state
is given by:
\begin{equation}
\label{E_corpa}
E_{gs}
= - \frac{\varepsilon N}{2}
  + \frac{1}{2} ( \omega - \varepsilon )
.
\end{equation}
%
%
\subsection{Generator coordinate method}

\subsubsection{Continous GCM}

Most applications of the GCM to the LMG model  are restricted to a
study of the GCM ground state and test the correlations in the
ground-state~\cite{Par68a}. In such cases, one can take the
``deformation'' $\alpha$ as a single generator coordinate. We are
here also interested in the description of excited states and will
also introduce $\varphi$ as a generator coordinate.

One can write the $N$-particle GCM wave functions as a linear
combination of the constrained HF states $|\alpha ,\varphi \rangle$
(\ref{HF:thouless}) with an unknown weight function $
f_{k}\left( \alpha ,\varphi\right)$
\begin{equation}
\label{eq:GCM:cont}
|\Psi _{k}\rangle
= \int_{-\frac{\pi }{2}}^{\frac{\pi}{2}} d \alpha
  \int_{-\frac{\pi }{2}}^{\frac{\pi}{2}} d \varphi \;
  f_{k} ( \alpha ,\varphi ) \; |\alpha ,\varphi \rangle
.
\end{equation}
Variation of the energy yields the so-called Hill-Wheeler-Griffin (HWG)
equation \cite{Hil53a}, an integral equation for $f_k (\alpha ,\varphi)$
\begin{widetext}
\begin{equation}
\label{hweq}
\int_{-\frac{\pi }{2}}^{\frac{\pi}{2}} d\alpha
\int_{-\frac{\pi }{2}}^{\frac{\pi}{2}} d\varphi \;
\left( \langle\alpha ^{\prime },\varphi ^{\prime }| \hat{H}
       |\alpha ,\varphi \rangle
       - E_{k} \langle \alpha ^{\prime },\varphi ^{\prime }
               |\alpha ,\varphi \rangle \right)\,
f_{k}\left( \alpha ,\varphi \right)
= 0
.
\end{equation}
The two kernels entering Eq.\ (\ref{hweq}) are the norm kernel
\begin{equation}
\langle\alpha^{\prime},\varphi ^{\prime }|\alpha ,\varphi \rangle
= \left[ \cos (\alpha) \cos (\alpha^{\prime})
        + \sin (\alpha) \sin (\alpha^{\prime})
          e^{ i ( \varphi - \varphi ^{\prime} )}
  \right]^{N}
,
\end{equation}
and the Hamiltonian kernel
\begin{eqnarray*}
\langle \alpha^{\prime },\varphi^{\prime }| \hat{H} |\alpha ,\varphi \rangle
& = & -\frac{\varepsilon N}{2}
      \langle \alpha ^{\prime },\varphi ^{\prime} | \alpha ,\varphi \rangle
      \nonumber \\
&   & \times
      \frac{ \cos^2 (\alpha) \cos^2 (\alpha^{\prime})
                   -\sin^2 (\alpha) \sin^2 (\alpha^{\prime})
                    e^{ 2i ( \varphi -\varphi ^{\prime } ) }
                   +\chi \left[ \sin^2 (\alpha) \cos^2 (\alpha^{\prime})
                                e^{ 2i\varphi}
                              + \cos^2 (\alpha) \sin^2 (\alpha^{\prime})
                      e^{-2i\varphi ^{\prime }}\right]
                  }{
           \left[ \cos (\alpha) \cos (\alpha^{\prime})
                + \sin (\alpha) \sin (\alpha^{\prime}) \;
                  e^{ i ( \varphi -\varphi^{\prime})}
              \right]^{2}
           }
.
\end{eqnarray*}
\end{widetext}
A set of orthonormal collective wave functions are obtained by an integral
transformation of $f_k$
\begin{equation}
\label{eq:wf_coll}
G_{k} ( \alpha^{\prime},\varphi^{\prime} )
= \int_{-\frac{\pi}{2}}^{\frac{\pi}{2}} d\alpha
  \int_{-\frac{\pi}{2}}^{\frac{\pi}{2}} d\varphi \,
  \langle \alpha^{\prime},\varphi ^{\prime }
          | \alpha ,\varphi \rangle^{\frac{1}{2}}
   f_{k} (\alpha ,\varphi)
.
\end{equation}
The HWG integral equation can be solved exactly \cite{RingSchuck}.
Thanks to the simplicity of the LMG model, the GCM with a single
generator coordinate $\alpha$ gives already the exact solutions of
the model. The GCM is build on signature-symmetry breaking constrained
HF states; hence, \emph{a priori} the GCM wave function mixes 0p-0h,
1p-1h, 2p-2h, 3p-3h etc states. However, one can easily see that the
signature symmetry is restored by mixing of HF wave functions
corresponding to $\pm \alpha$ with weights equal in modulus, which comes out
automatically from Eqn.~(\ref{eq:GCM:cont}). This situation is
similar to realistic applications of the GCM when a discrete
symmetry, like parity, is broken at the mean-field level but not for
continuous symmetries like rotations.

%
%
\subsubsection{Discretized GCM}
In realistic calculations \cite{Val00a,BH03,Rod00a,Nik06a}, the HWG
equation is solved by discretization of the collective variables
\begin{equation}
\label{eq:GCM:disc}
|\Psi _{k}\rangle
= \sum_{\alpha} \sum_{\varphi}
  f_{k} ( \alpha ,\varphi ) \; |\alpha ,\varphi \rangle
.
\end{equation}
The discretized GCM equations are obtained by replacing all
integrals in Eqns.\ (\ref{eq:GCM:cont}-\ref{eq:wf_coll}) by sums
over discretization points. The integral equation (\ref{hweq})
becomes a matrix equation which can be solved by diagonalization.

As our aim is to understand why the GCM overestimates excitation
energies in realistic applications, we will solve the HWG equation
of the LMG model by discretization. Since the continuous GCM
permits to find the exact eigenstates of the LMG Hamiltonian, the
number of discretization points must be small enough to avoid a
trivial reproduction of the exact solution.

\begin{figure*}[t!]
\includegraphics[width=16cm]{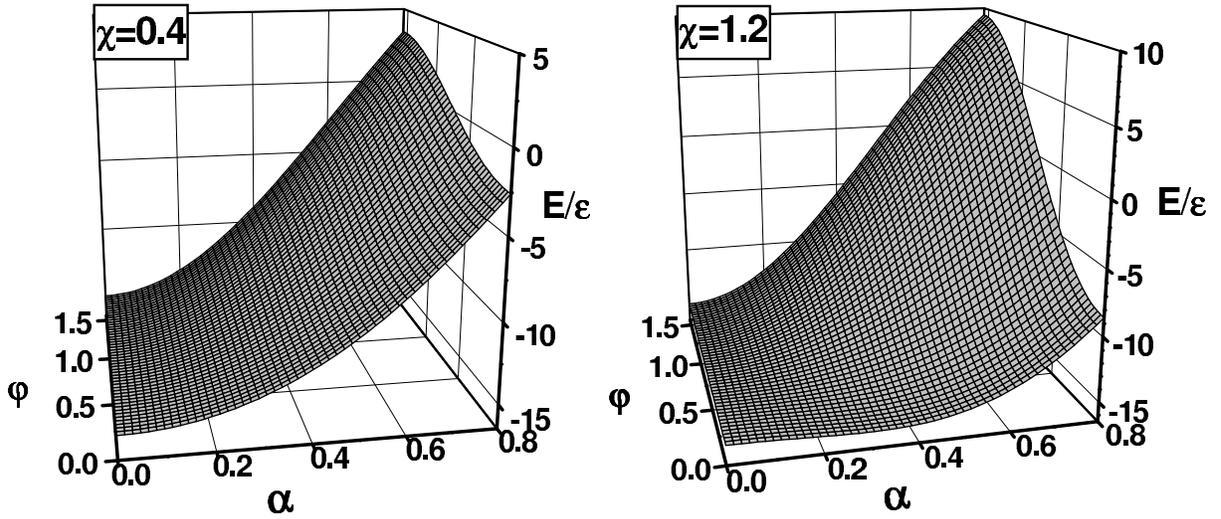}
\caption{\label{E_lmg}
Mean-field potentials for two values of the interaction strength,
\mbox{$\chi=0.4$} and \mbox{$\chi=1.2$}, for \mbox{$N=30$} particles.}
\end{figure*}

We have solved the LMG model for 30 and 50 particles, with a
discretization on meshes symmetric around $0$ in $\alpha$ and
$\varphi$ and with an odd number of points. The meshes in $\alpha$
and $\varphi$ have been limited to the region in which the
collective wave function has a sizable amplitude, typically less
than half the total range of variation of $\alpha$. The mean-field
potentials that are obtained for two values of $\chi$, above and
below the phase transition, are plotted in Fig.~\ref{E_lmg}. The
mean-field ground state corresponds to $\alpha$ and $\varphi$
equal to zero when $\chi$ is smaller than 1, and to a non-zero
value of $\alpha$ otherwise. The energy surface is very flat for
small values of $\varphi$; it is only when $\alpha$ is large that
the energy increases rapidly with $\varphi$. Such a topography is
representative of the deformation energy curve that is obtained
for spherical nuclei (\mbox{$\chi < 1$}) and for nuclei soft as a
function of deformation (\mbox{$\chi > 1$}).

\begin{figure}[t!]
\includegraphics[height=8cm]{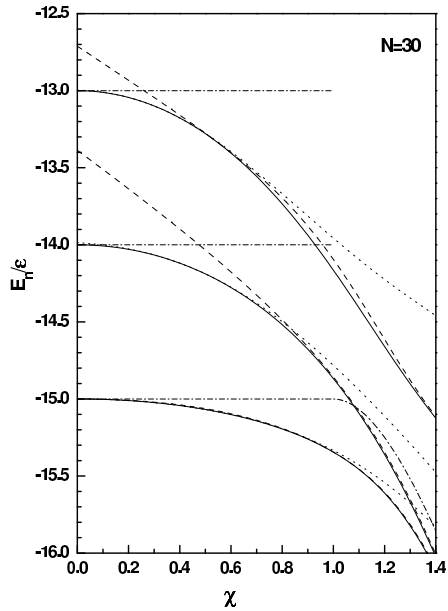}
\caption{\label{e_gcm1}
Comparison between the exact and GCM energies of the first three
states of the LMG model for 30 particles. The solid lines are
the exact result, while the dashed and dotted lines represent the
GCM solutions obtained with the two different meshes (see text)
with seven points in $\alpha$. The HF results are indicated by a
dash-dotted line.}
\end{figure}

In Fig.~\ref{e_gcm1}, the exact and GCM energies for the  ground
state and for the first two excited states are plotted as a
function of the two-body interaction strength $\chi$ for a system
of 30 particles. We have used $\alpha$ as the only  generator
coordinate. The GCM equations are solved with 7 equidistant
discretization points chosen in the region where the collective
wave functions have a sizeable amplitude. In realistic
applications of the GCM, the overlap kernel is used to define the
mesh, the requirement being that the overlap between two adjacent
points is of the order of 0.8. However, in the LMG model, this
kernel does not depend on the two-body interaction strength. Since
we want to be as close as possible to realistic applications,
where the exact solutions are not known, we have taken the same
mesh for all interaction strengths. From the collective wave
function obtained for a value of $\chi$ around 0.5, we have chosen
points in the interval ($-0.707$, $+0.707$). Since the wave functions
of the model are either even or odd with respect to $\alpha$, only
four discretization points are significant.

For \mbox{$\chi < 1$}, the HF ground-state energy does not depend
on $\chi$, as the ground state is always "spherical".
The correlations beyond mean-field significantly improve the HF
result for the ground state and bring it very close to the exact
value for all values of the interaction strength. The situation is
less satisfactory for excited states: in particular, their excitation
energies are far from the exact values when the interaction is switched
off, while the HF method gives the exact energies. The discretized GCM
becomes more accurate than HF only for $\chi$ larger than 0.25 for the
second excited state and than 0.45 for the first one.

To understand this surprising result, let us analyse in more
details how the continuous GCM works in the \mbox{$\chi = 0$}
limit. Let us first note that, while the exact ground state wave
function, corresponding to $\alpha$ and $\varphi$ equal to zero,
is included in the generating functions $|\alpha ,\varphi
\rangle$, the first excited states corresponding to pure 1p-1h and
2p-2h excitations are not. One can easily verify that the weight
functions that permit to extract the exact eigenstates are the
Dirac $\delta(\alpha)$ distribution for the ground state, its
first derivative for the first excited state and combination of
$\delta(\alpha)$ and its derivatives for higher excitations. The
collective wave functions, given by equation~(\ref{eq:wf_coll}),
are regular functions but with rather sharp peaks. One of them is
located at $\alpha$ equal to 0 for even signature states, while
the odd signature states vanish at 0. Moreover, for very small
values of $\chi$, the collective wave-functions have very small
amplitudes at the most extreme mesh points, leaving only a very
small number of significant discretization points.

Replacing one of the mesh points by a point close to the extrema
of the wave functions of the first and second excited state for
$\chi$ equal to zero, the discretized GCM results become very
close to the exact values for the three first states and for low
interaction strengths. For the first excited state, which has a
node at the origin, this discretization is superior to the one
based on the \mbox{$\chi = 0.5$} ground state wave function up to
\mbox{$\chi=0.75$}.

\begin{figure}[t!]
\includegraphics[height=8cm]{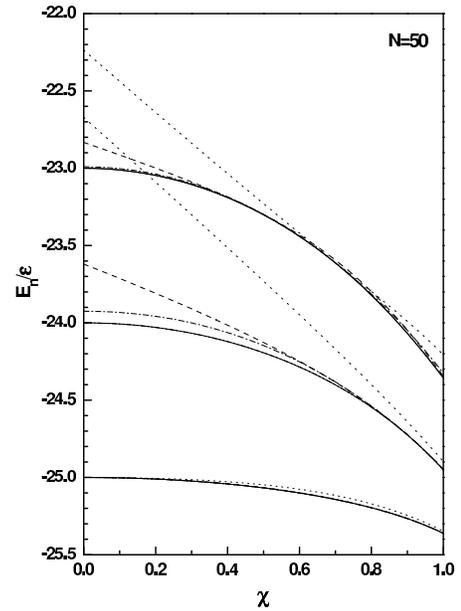}
\caption{\label{e_gcm2}
Comparison between the exact and GCM energies of the first three
states of the LMG model for a system with \mbox{$N=50$}. Two generator
coordinates, $\alpha$ and $\varphi$ are used in the GCM. The exact
result are represented by solid lines. The dotted and dashed lines
lines are GCM results with 7 and 9 points in $\alpha$; the
dash-dotted lines are results of calculations with 7 points in
$\alpha$ and 3 points in $\varphi$.
}
\end{figure}

%
%
\subsubsection{The role of the second generator coordinate}
The influence of a second generator coordinate is shown in
figure~\ref{e_gcm2}, where are plotted the energies of the first
three states of a system with 50 particles. The calculations are
performed with 7 and 9 points in $\alpha$, and 1 or 3 points in
$\varphi$. For the mesh in $\alpha$ we choose equidistant points 
in the interval (\mbox{$-0.660$, $+0.660$}).
Since the mesh in $\varphi$ is symmetric and
\mbox{$\varphi=0$} is always a mesh point, there is in practice
one active point in $\varphi$. All discretizations give accurate
results for the ground state. With a mesh in $\alpha$ only, the
energies of the first excited state are inaccurate, for all
interaction strengths with 7 points, and below $\chi$ equal to 0.4
for a 9 point discretization. The second excited state is slightly
better described although the accuracy is still limited for small
interaction strengths. Adding points in $\varphi$ to the calculation
with 7 points in $\alpha$ corrects the behavior near the origin and leads
to very accurate results for the ground state and the second excited
state. The energy of the first excited state is also improved
although there still remains a small discrepancy for small
interaction strengths. The combination of nine points in $\alpha$
and three points in $\varphi$ leads to results indistinguishable
from the exact ones.

\begin{figure}[t!]
\includegraphics[height=8cm]{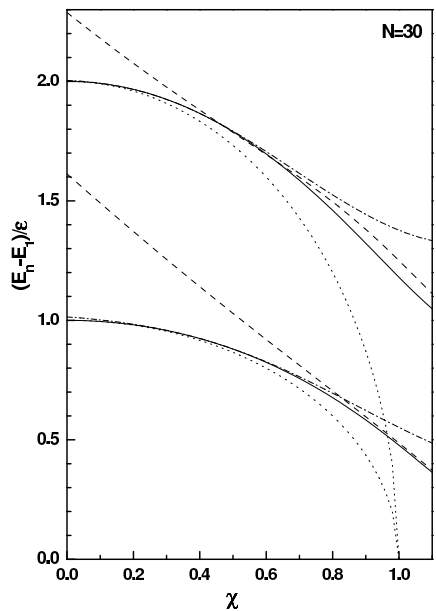}
\caption{\label{E_excit}
Excitation energies of the first two excited states as a function
 of the interaction strength $\chi$. Exact results are given by solid lines,
 the RPA results by dotted lines. The dashed and
dash-dotted lines correspond to GCM calculations with the same
choices of points as in figure~\ref{e_gcm1}.
}
\end{figure}

\subsubsection{Comparison between the GCM and the RPA}

The RPA permits to determine excited states but also correlation
energies in the ground state. The correlations that are given by
Eqn.~(\ref{E_corpa}) are very accurate and makes the RPA ground
state energies very close to the exact values up to interaction
strength equal to 0.9. Beyond this value, the GCM with 7
discretization points is more accurate than the RPA.

The GCM and RPA results for the excitation energies of the first
two excited states are compared to the exact values in
Fig.~\ref{E_excit}. It is the accuracy of this energy difference
which is the most interesting in usual applications of the GCM.
The use of a mesh adjusted for an intermediate interaction
strength (\mbox{$\chi = 0.5$}) and on the ground state gives
values for the energies of the second excited state close to the
exact ones when the interaction strength is large but it fails for
weak interactions. The situation is even worse for the first
excited state which does not have the same symmetry as the ground
state. In this case, results are close to the exact ones only
beyond $\chi$ equal to 0.75, the error being as large as 60$\%$
for $\chi$ close to zero. The results are by far better when the
mesh is adapted for low-$\chi$ values, as discussed above. Then,
even for a small number of mesh points, the exact results for both
excited states are correctly reproduced by a discretized GCM
calculation. The RPA results have a very different behavior. They
reproduce the exact results quite well for small interaction
strengths. The first excited state becomes inaccurate above $\chi$
equal to 0.7; the second one deteriorates more quickly and is
worse than the blind GCM discretization above $\chi$ around 0.5.

%
%

\section{Conclusions}

In this paper, we have analyzed possible causes of the inaccuracy
on energies of excited states that are predicted in GCM
calculations. The study of Sn isotopes has shown that part of the
overestimation of excitation energies can be due to the effective
interaction: RPA results obtained with the same Skyrme interaction
for (\nuc{132}{Sn}) are also in disagreement with the experimental
data. Although it is not easy to perform fully equivalent RPA and
GCM calculations, it seems clear that the GCM fails to reproduce
RPA results for spherical nuclei and that at least part of the
energy overestimation obtained in previous works is due to the way
we use the GCM.

The use of a schematic model does not allow to draw final
conclusions on the origin of the problems encountered in realistic
GCM applications, but it can give some hints on it. To the best of
our knowledge, we have analyzed for the first time the application
of the discretized GCM to the LMG model. To summarize our results,
one can say first that the discretized version of the GCM works
remarkably well and permits to reproduce the exact results with a
very limited number of points. Of course, the dimension of the LMG
model is very limited but nearly exact results are obtained for
the three first states using only approximately $1\%$ of the total
number of independent vectors of the LMG space. A second result is
that correlations in the ground state are better described than
correlations in excited state. We have also seen that results are
closer for the second excited state which has the same symmetry as
the ground state than for the first one. The fact that an
appropriate choice of a small number of discretization points
permit to obtain excellent results, better than the RPA, seems to
be an artefact of the LMG model, hard to transpose on realistic
cases. On the other hand, it seems encouraging that the excited
states are well described by the GCM when the collectivity due to
the generator coordinate is large. Finally, the introduction of a
second generator coordinate, conjugate to the first one, also
improves the GCM results and seem to make them more independent on
the way the discretization is performed.

How to transpose the LMG results on realistic cases is not
trivial. The fact that the LMG mean-field states break all
symmetries of the interaction certainly reinforces the assumption
that to break more symmetries of the nuclear Hamiltonian will
improve the GCM description of excited states, in particular by
introducing states breaking time reversal invariance. This could
be done in several ways. The most sophisticate one would be to
introduce collective variables conjugate to the deformation modes
of the nucleus, as in the adiabatic time-dependent HF formalism of
Villars \cite{Vil77a}, Goeke and Reinhard \cite{Rei78a}, or of Baranger
and V{\'e}n{\'e}roni \cite{Bar78a}. A more economic way to proceed
along similar lines is to introduce a few states breaking time-reversal
invariance which are guessed to enlarge strongly the variational
space: cranking constraint or specific two-quasi-particle excitations.
Work along these lines is in progress.

%
%

\section*{Acknowledgments}

We thank G.~Col\`o for providing us with the result of the QRPA
calculation and G.~F.\ Bertsch, W.~Nazarewicz and H.~Flocard for
inspiring discussions.
This work was supported in parts by the Belgian Science Policy Office
under contract PAI P5-07 and by the U.S.\ National Science Foundation
under grant no.\ PHY-0456903.
%
%

\end{document}